# Visual Integration of Data and Model Space in Ensemble Learning

Bruno Schneider*   Dominik Jäckle[†]   Florian Stoffel[‡]   Alexandra Diehl[§]   Johannes Fuchs[¶]
Daniel Keim[‖]

University of Konstanz, Germany

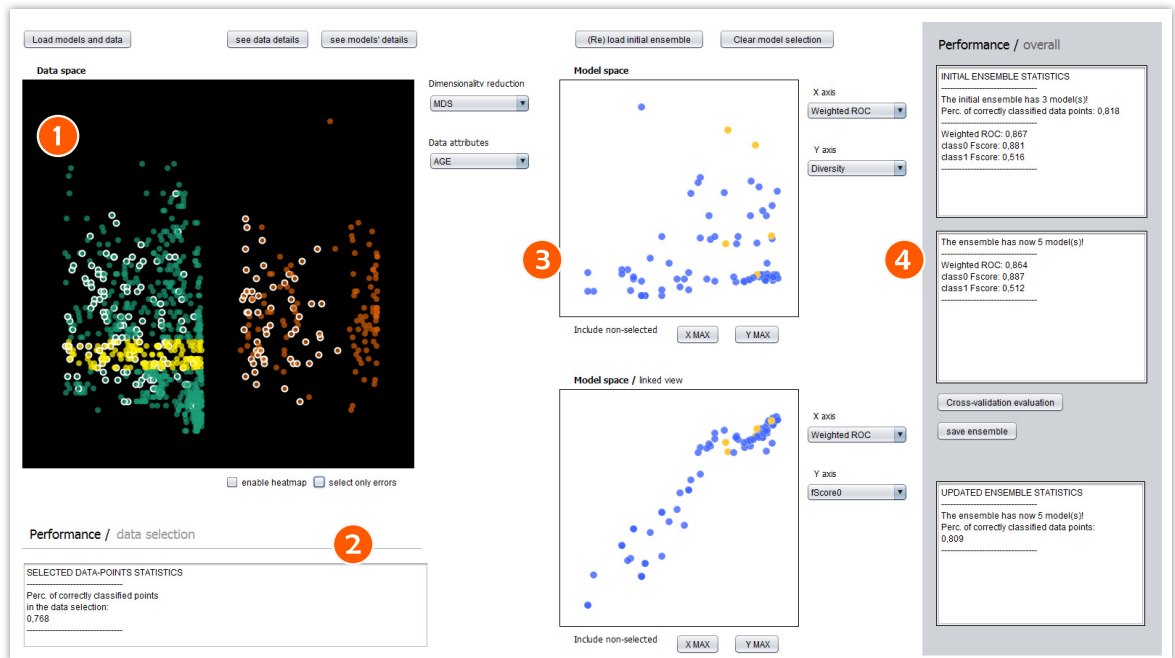

Figure 1: **Overview of our tool for exploring ensembles of classifiers**. The user can select regions of the classification outputs (1), see how the ensemble classified these regions (2), interact with a preloaded collection of models adding or removing them to update the ensemble (3), and track the performance while interacting with the models (4).

**ABSTRACT**

Ensembles of classifier models typically deliver superior performance and can outperform single classifier models given a dataset and classification task at hand. However, the gain in performance comes together with the lack in comprehensibility, posing a challenge to understand how each model affects the classification outputs and where the errors come from. We propose a tight visual integration of the data and the model space for exploring and combining classifier models. We introduce a workflow that builds upon the visual integration and enables the effective exploration of classification outputs and models. We then present a use case in which we start with an ensemble automatically selected by a standard ensemble selection algorithm, and show how we can manipulate models and alternative combinations.

**Index Terms:** Pattern Recognition [I.5.2]: Design Methodology—Classifier design and evaluation

*e-mail: bruno.schneider@uni-konstanz.de
[†]e-mail: dominik.jaeckle@uni-konstanz.de
[‡]e-mail: florian.stoffel@uni-konstanz.de
[§]e-mail: diehl@dbvis.inf.uni-konstanz.de
[¶]e-mail: fuchs@dbvis.inf.uni-konstanz.de
[‖]e-mail: keim@uni-konstanz.de

## 1 INTRODUCTION

Given a set of known categories (classes), *Classification* is defined as the process of identifying to which category a new observation belongs. In the context of machine learning, classification is performed on the basis of a training set that contains observations whose categories are known. A key challenge in classification is to improve the performance of the classifiers, hence new observations are correctly assigned to a category. Classification can be performed with a variety of different methods tailored to the data or the task at hand. Examples include, among others, decision trees, support vector machines, or neural networks. Research proposes to improve the accuracy of classification using *Ensemble Learning* [11, 36], also known as *Multiple Classifier Systems* (MCS) [31]. Such systems suggest to combine different classifiers, each targeting a different task.

Well-known approaches for building ensembles propose to either train the same model successively with different subsets of the data [4, 12], to combine different model types [18, 34], or to combine different strategies such as bagging [4] with random feature combinations in Random Forests [5]. Generally speaking, the application of ensembles increases the complexity of the classification process bringing in the inherent problem of decreasing comprehensibility. In particular, it is challenging to understand how and to what extent the models contribute to the classification, as well as which models produce a significant number of classification errors.

Visual and automatic methods for the analysis of Classification outputs in Ensemble Learning typically do not provide a direct link from the data space back to classification model spaces with other candidates for experimenting with new ensemble configurations. Regarding the visual methods, they also often do not scale properly to represent a greater number of classifiers in ensemble model spaces. For example, in [32] Silva and Ribeiro show how the models contribute individually, but the analysis is limited to inspect the ensemble after making the decision of which models will take part on it. In [33], Talbot et al. present a system in which is possible to interact and combine models and their classification outputs through *confusion matrices*, but with a limited set of model candidates. However, if we connect the data space with representations of a big set of different classifiers that covers a wide range of the model parameter space, we foster an analysis process with a feedback loop that allows successive local improvements that are not given by any visual or automatic method for analyzing and exploring ensembles. In this work, we aim to address the research question: *How to integrate data and model space to enable visual analysis of classification results in terms of errors in Ensemble Learning?*

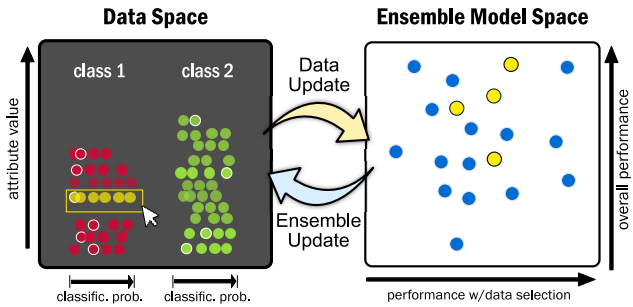

Figure 2: Visual integration of the data and ensemble model space. Left: The classification results are binned per attribute and class. The visual representation aligns the data points regarding their attribute value and the classification probability. A manual selection in the data space triggers a *data selection update*. Right: The model space depicts each single model and allows to compare them by customizing the axes; herein, we contrast the overall performance with the performance w/ data selection. The interactions in the model space trigger an *ensemble update* with immediate impact on the data space.

We propose an interactive visual approach for the exploration of classification results (data space) in close integration with the model space. Its main goal is to give direct access to models in classifier ensembles, thus enabling to experiment with alternative configurations and seek for local classification patterns that are not visible through aggregate measures. We visualize each classified data point and then provide direct access to each individual model that is part of the ensemble. Figure 2 depicts our approach. The data points are binned per class and attribute. For each bin, the points are visually aligned regarding the attribute value and the classification probability, which enables the identification of local areas of classification errors and areas of high classification certainty or uncertainty, respectively. In the current iteration, our implementation is focused on data scientists with good knowledge of the data and models at hand.

In this work, we claim the following two-fold contribution towards enabling the visual analysis of classification results in Ensemble Learning: First, *the tight visual integration of the data and the model space*. Second, *a workflow that builds upon the visual integration and enables the effective exploration of models and classification outputs*. The visual integration allows to manipulate and explore the impact of each data object and model in a straightforward manner. We provide visual guidance to identify effective models not selected by the automatic search in first place. One can then include the identified models into the ensemble for local improvements based on the constraint that the overall performance is not impaired. The views update on the fly, enabling the user to retrace the impact on the classification output.

## 2 RELATED WORK

Our work builds upon the idea of visually integrating the space of *machine learning* models and the data space, thus enabling the exploration of the impact of each data object and model. Following, we discuss related work from ensemble learning and interactive model space visualization. Our approach does not aim at retraining the models but at finding effective model combinations that were not given by the automatic search. Therefore, we do not discuss the family of well-known visualization methods with respect to the data space.

### 2.1 Ensemble Learning

Classifier ensembles aim at combining the strengths of each classification model. To build ensembles, it is necessary to generate a variety of models and then to combine their results. The first step – generating the diversity of models – can be accomplished by making use of different strategies. Several ensemble learning philosophies [15] and methods for combining the classification outputs [28] exist. For example, the same model can be trained successively with different subsets of the data [4, 12], with different types of models [18, 34] (e.g. Decision trees, K-nearest neighbors), or with combinations of strategies such as the mixture of bagging [4] and random combinations of strategies in the Random Forests [5]. In our case, we generated diversity by producing distinct types of model. The main reason for this choice was because the model diversity produced by the other strategies is often given by the design of the respective algorithms. In these cases, it is only necessary to set a base classifier and all the other models are automatically generated in the background (e.g. the AdaBoost M1 method [12], in which usually Decision Stump trees are used as the base classifier to produce ensembles using a boosting strategy). Conversely, with multiple types of classifiers it is necessary to define each model that will take part in the ensemble, and this selection procedure and the multitude of possible combinations motivated the use of data visualization to support the selection task. In particular, we worked with Multiple Classifier Systems, in which there is usually an overproduction phase and the generation of big model libraries (with hundreds or even thousands of models because the analyst typically does not know beforehand which model types will perform well together). Then, with the big model libraries, there are several search algorithms that were developed to look for the best possible combination of models automatically (e.g. GRASP [23, 35], evolutionary algorithms [1]), without experimenting with all the possible combinations due to the complexity of this combinatorial problem. In our work, we used a search selection algorithm developed by Caruana et al. [9] that also implement strategies to avoid over-fitting [8], a common issue with ensembles of classifiers.

### 2.2 Interactive Model Space Visualization

Following, we provide an overview of visualization techniques to represent the model space or also called *model landscape*. Building upon the well-known visualization methods, we then discuss interactive approaches introduced to steer the performance of ensemble classifiers.

Rieck et al. [30] used scatter plots for representing regression models, in particular, and to perform a comparative analysis of competing regression models. In contrast, Olah [26] also shows groups of models using scatter plots, but for representing distinct Neural Network architectures to classify images of hand-written digits. In a similar way, Padua et al. [27] represented collections of *Decision Trees* using several linked visualizations, in which the users can explore large portions of the parameter space of these models and assess the predictive quality of trees derived from several combinations of parameters. A relevant aspect of building visual representations of models and doing it in a scalable way is that we can link them to the data. Visualization, in synergy with interaction, can play an important role for this direct link between data and models [3].

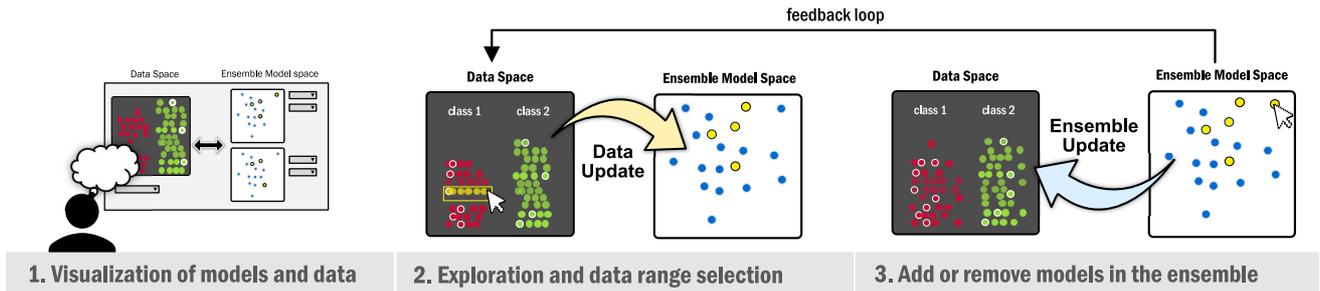

Figure 3: Our process for exploring ensembles of classifiers starts with the visualization of the classification outputs and the combination of models that produced the corresponding classification (1), allows the user to explore and select regions of interest in the data space (2), updates the model space to show how each individual model classifies the current data selection, and allow inclusion or removal of models from the ensemble (3). Then, at each change in the ensemble configuration, the visualization of the classification outputs is updated accordingly, which introduces a feedback loop that can lead to new rounds of interaction with the system.

The analysis of classifier models through interactive visual interfaces is an active area of research [22]. Talbot et al. [33] present *EnsembleMatrix*, in which the user can interactively build and steer the performance of ensembles of classifiers. In [16], Kapoor et al. also present an interactive tool called *ManiMatrix*, in this case for the improvement of individual classifiers. In both cases, *confusion matrices* appear as a central component. In *ManiMatrix*, the users can express their preferences w.r.t. decision boundaries among classes using a confusion-matrix. In *EnsembleMatrix*, the matrices support the decision of which combination of classifiers works better when building ensembles. However, despite the compact and efficient information about the class confusion that these matrices convey it is still aggregated data about performance that does not go until the bottom level of the errors with individual data points. To provide this level of access and better visualize where are the errors coming from, we worked with a representation of the data space that shows this level of detail. Also, due to scalability issues the decision to show one classification matrix for every classifier in *EnsembleMatrix* is not applicable to our case, in which we had libraries with hundreds of models for building MCS. In *EnsembleMatrix*, the ensembles were built from a limited and small number of candidate models, and not in the same way that happens in our context of building MCS. Regarding giving access to the data instead of only showing aggregated information about model errors, Ren et al. [29] pointed recently this limitation of most current systems. They presented a solution for multi-classification problems in which they visually compare different models with similar performance but with very distinct behavior w.r.t. to the classes and local regions of the data space. *ModelTracker* [2] also provided access to the data level for model performance analysis. We go in the same direction of revealing errors that are not visible when aggregated but we do that in an Ensemble Learning context. While our visual interactive approach supports improving classifier models, we particularly focus on the integration between classification results and models and propose a workflow for the effective analysis.

## 3 INTEGRATING DATA AND CLASSIFICATION MODEL SPACES

We propose a visual analytics approach for the exploration of model spaces in ensembles of classifiers. We work with Multiple Classifier Systems (MCS) and introduce a data-guided and user-oriented process for interacting with data and models in this context (see Figure 3). In addition, the direct linkage of data and models is a central component of our workflow, because it allows the user to manipulate objects in any side and see the impacts on the other side instantly, by means of interaction and data visualization.

MCS are often generated from huge model libraries of several types of classifiers with different parameter settings. The process does not depend on previous knowledge about which models perform better for the data and classification task at hand. Several models are produced and an automatic search step looks for the best possible combination of models that deliver higher performance when combined in an ensemble of classifiers. In our workflow, we build a MCS using the standard automatic approach previously described. Then, we initialize our tool, in which we can visualize the models and the classification outputs produced by the initial automatically selected ensemble. Our starting point with our visual analytics approach is after the automatic construction of MCS.

In our tool, we have a visualization panel that represents the classification outputs (the data space, see Figure 1 (1)), and two linked other ones that show the classification models accordingly to selectable performance and diversity measures (the model space, Figure 1 (3)). Importantly, we show not only the models that were automatically selected and correspond to the initial ensemble configuration, but also show the whole model library that was used in the beginning of the construction. Therefore, it is possible to add or remove models at any time to the ensemble. With our approach, the process of exploring the model space is driven by the user interest in particular regions of the data space. We present our model space exploration process and its feedback loop with greater detail in the next subsections.

### 3.1 Representing models and data

We aim at enabling the user to directly manipulate each data point and each model in our visualizations. With respect to scalability, we have to consider that the model libraries for building MCS can have hundreds of classifiers. Regarding the data space, we always visualize the test dataset with unseen data during the training phase of the models. In any case, models or data, we need a visualization that can accommodate these objects at scale. For this reason, we decided to use scatter plots to visualize both. In the model space, each dot corresponds to a classifier model and the color shows if the model is part of the current selected ensemble or not. Analogously, in the data space each dot corresponds to one data item of the test data set and the color indicates the predicted label.

In our tool (Figure 1), the left-side panel is the data space. The user can select one attribute at each time (e.g. *age*) and visualize how the data is classified by the ensemble w.r.t. this attribute. This representation is relevant for our approach because it can reveal interesting regions of the data space in the attribute level and point to models and their performance for these particular selections. In this visualization, besides taking one of the dimensions of the scatter plot and representing the attribute values on the vertical axis we also compute the classification probabilities for each data point of the predicted class and map to the horizontal axis. In addition, we display each class in a different region of the plot and side-by-side, to better distinguish the classification outputs per class (see Figure 5, in which we have a binary classification problem and the green and orange colors distinguish the data instances from both classes.). However, in scatter plots the overplotting can occur and make it difficult to better identify dense regions in the data. To overcome this problem, we implemented and included a heat map visualization in the data space. At any time, the user can switch between the standard scatter plot and the heat map to show the same data. With this heat map, the clusters with classification errors become more distinguishable. Also, the user can decide to have a first impression of the classification outputs by selecting to project the data attributes to the two-dimensional space using *PCA*, *MDS* or *t-SNE* methods.

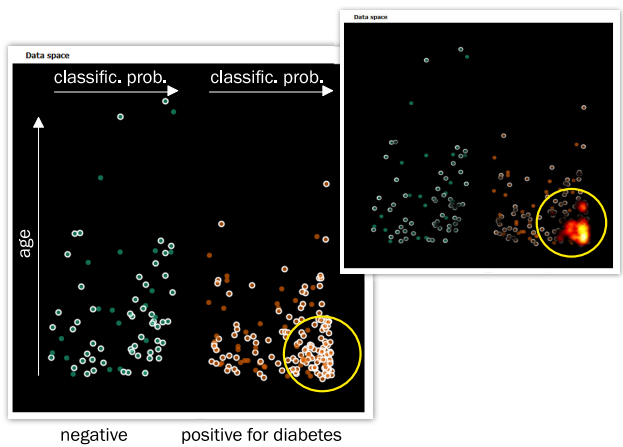
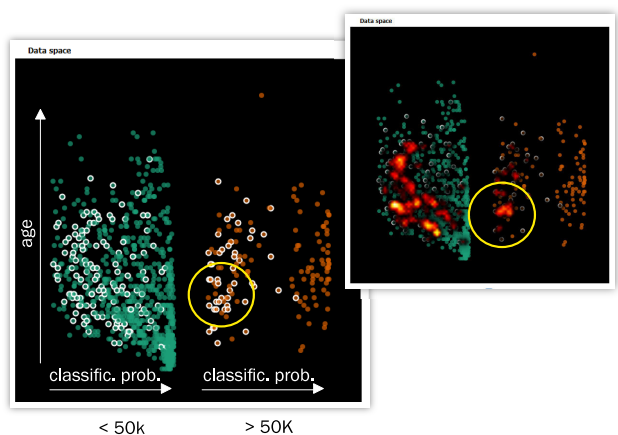

Figure 4: On the left image, we see the classification outputs for a dataset of persons that tested negative or positive for diabetes. Circled in yellow, there is a cluster of mislabeled data points in the *positive* class (dots with a white outline) that corresponds to individuals with lower age. In these cases, the ensemble assigned high classification probability, which makes these errors more difficult to fix. The right image shows the same data with a heat map on top of it.

Figure 5: On the left image, we see the classification outputs for a dataset of persons that earned less or more than 50K per year. Circled in yellow, there is a concentration of mislabeled data points in the *more than 50k* class (dots with a white outline). In these cases, the ensemble assigned low classification probability, which makes these errors more easy to fix. The right image shows the same data with a heat map on top of it.

With respect to the models, we represent them in our tool in the right-side scatter plots (Figure 1). We precompute measures of performance and diversity for each model and let the user decide which one should be assigned to each axis. We have two linked panels for the models because this layout gives more flexibility to the user and allows the visualization of the same group of models from different perspectives in each of the panels. For overall performance, we have the weighted *Area Under ROC* score, and the user can also choose the *F-Measure* score per class. Regarding the diversity, we use the *Q-statistics* [19], a pair-wise measure that compares the classification of each data point between two classifiers and captures if the models classify the data in a similar way or not.

In the context of ensemble learning, model diversity is an important aspect. Very often, we want to find models that classify distinct regions of the data space differently, because then we can combine the model strengths in a good ensemble. There is research about the role of diversity in ensembles [7] and it is not guaranteed that we can always use this measure to get the best model combination. However, it is still a relevant metric to consider when comparing and visualizing classifiers in ensemble model spaces. With the diversity measure, we can read the plots in the following way: models more close to each other in any of the axis (similar *Q-statistics* value) classify the data in a similar way, and models far away from each other classify the data differently, despite the fact that they can still have similar performance. Lastly, we obtained the individual measure of diversity for each model by taking a square matrix with the *Q-statistics* for each pair of models, applying PCA and then taking the first component of it.

### 3.2 Interacting with data and model representations

The model-data linkage is a central component of our approach. With this link, we can have the visualization of the data space as an entry point for the user to find regions of interest in the data and the corresponding performance of the models for these regions. This behavior is backed, naturally, by a series of interactions that we implemented in both model and data panel visualizations. We have, at the end, a process that contains a feedback loop (Figure 3), in which for any data selection we have corresponding model candidates, and them for any model selection the classification outputs in the data space change again, potentially allowing new rounds of interaction.

When the user interacts with the data space and selects items, we compute the performance of each individual model for the current data selection (percentage of correctly classified data items). Then, the user can decide to show in one axis of the model scatter plots this local performance, and the model space will update accordingly to the selection. In addition, it is also possible to activate a filter that selects only the points that were wrongly classified by the ensemble. This functionality makes it easier to take care of the errors in separate, by facilitating a fast selection of these data points. The data space always represents the classification outputs (predicted class, corresponding classification probability, and misclassified items) for the current ensemble selection in the model space visualization.

The user can also interact with the models. In this case, when a model is selected we automatically update the ensemble that now has one more model, and we compute the classification outputs for the new ensemble selection. The accuracy of the ensemble is updated at each interaction with the models and displayed in a text panel with the percentage of correctly classified points in the test dataset. This computation is very fast to do because we already have precomputed the results for each available classifier in the model space, so it is only necessary to combine the results of the selected models at each time the selection changes. We use the arithmetic mean of probabilities method, which is a standard procedure to combine classification results in an ensemble (for more details and other possible methods, see [20]). However, to get more stable results about the ensemble performance, we also have a button that the user can press to perform a cross-validation evaluation (set by default for 5-folds), which takes more time to compute and for that reason is not configured to respond automatically to any change in the model space. In addition, the user can also reinsert models, by just clicking again on already selected models. The data space is always updated at each model inclusion or removal in the ensemble, to show the classification outputs accordingly to the models' selection.

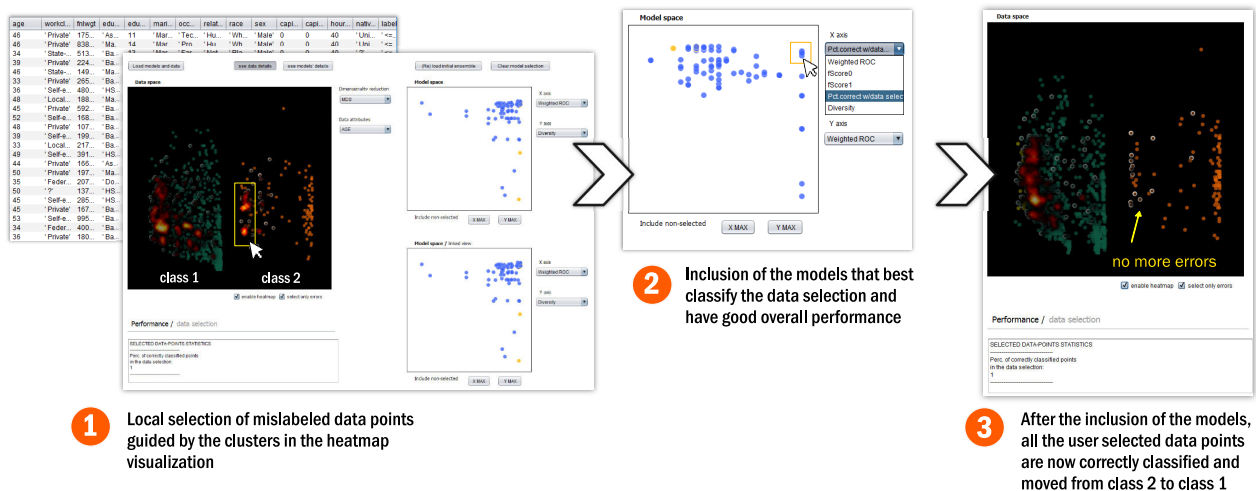

Figure 6: **Use case: visualizing clusters and improving the classification of user-defined regions in the data space**. We provide an example of how to use our approach for identifying interesting regions in the classification outputs (1), to select from the models available in a preloaded library the ones that are not part of the ensemble yet and perform well for the data selection (2), and finally for observing again in the data space how the classification changed (3) after adjusting the ensemble configuration and selecting models that were not considered by an automatic ensemble selection algorithm.

### 3.3 A data-guided and user-oriented exploration process of classification model spaces

In our tool, the user can freely interact with models or data, in any order. However, we designed our process to have the data space at the beginning of the interaction. For that, we gave more space to it in our graphical user interface. In particular, the addition of a heat map on top of the data space scatter plots facilitates the identification of potentially interesting clusters in the classification outputs. Our workflow and visual tight integration of data and models aim to provide an avenue for the exploration and comprehension of classification results in ensemble learning. Our approach can also be used for performance improvement, but this was not our main focus at this time.

We designed our data exploration process taking into account the possibility of filtering the data space by one attribute at each time, which potentially allows the user to do meaningful selections and identify particular regions of major interest. This gives to the user the chance to deeply explore the classification outputs, understand their relationship with the selected data attribute and identify clusters of errors that could not be distinguishable in overview visualizations (like the ones obtained with *PCA*, *MDS* or *t-SNE* methods applied to all data attributes).

Naturally, we can have datasets in which some attributes or dimensions have no meaning to the user. However, in this case it is still possible to find clusters of classification errors if they exist, and look for models and its corresponding performance with these data points. Also, despite the fact that our visualization of the classification outputs allows easy identification of clusters of errors per attribute, the process we introduce is a very exploratory one. For instance, the user can find models that are not part of the ensemble yet and perform well for the data selection while showing a good overall performance. In these situations, we have potential candidates for inclusion, and the user can naturally do that by just adding them and tracking the performance in the provided performance text panels. However, it can also happen that the models available for inclusion do not help with a particular selection of the data space. In these cases, our approach still plays the major role of enabling the exploration of data and models together, increasing the comprehensibility in ensemble learning. We see, then, our process as a data-guided and user-oriented exploration of ensemble model spaces that supports the discovery of interesting classifiers when they exist, given the data and classification task at hand.

## 4 USE CASE: VISUAL ANALYSIS OF CLASSIFICATION RESULTS IN A BENCHMARK DATASET

To showcase the usefulness of our approach, we built a Multiple Classifier System and applied it to a classification problem using a benchmark dataset. We selected the *Adult Data Set* [21], in which the classification task is to predict whether a person makes over fifty thousand US dollars per year. Then, we took an existing collection of classification models, available in [14], and corresponding parameter specifications to build MCS for binary classification problems. We prepared a subset of this reference library with 100 models and distinct learning strategies (12 model types), with varying parameter settings. Then, we built an automatically selected ensemble from this library of models using an implementation of the ensemble selection algorithm of Caruana et al. [9], parameterized to run using backward and forward search together, also available in [14]. The ensemble selection algorithm picked five models from the library to build our initial MCS (one bayesian network classifier, three bagging models – two J48 trees with distinct parameter settings and one REP Tree as the base classifiers in each case – and a random forest model). We trained the models with 32,500 data instances. Then, we selected a group of 1000 data points not seen during the training phase to show in our visualizations and interact with this subset. All the measures about individual model performance, which we need to represent them in our model space visualizations, were collected using a 5-fold cross-validation evaluation.

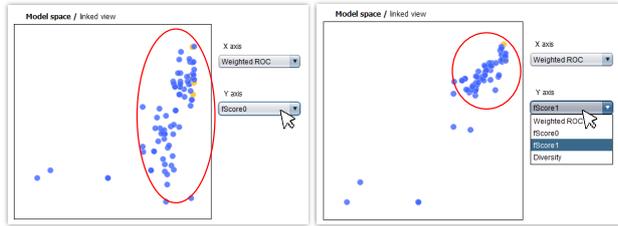

Figure 7: Exploration of the ensemble model space. Left: When the user adjusts the axes in the model space to show the classifiers with both good performance for the class 0 (less than 50k per year) and globally, we see a dispersion along the vertical axis. This means that not the majority of available models classify well the selected class. Right: Conversely, the availability of models to improve the other class is higher (cluster circled in red).

After setting up all that was necessary to run our experiment, we initialized our tool with the mentioned data and models. At the beginning, the system starts with the automatic selected models in the previous step. The user can interact with this ensemble, and also with all the others that were not automatically selected but were available in the collection of models that originated the MCS. With our tool, it is possible to freely experiment with any ensemble configuration and start the interaction with the data or with the models. However, in the following we describe our usage example in accordance with the workflow presented in Figure 3. The idea is starting by inspecting the data and finding clusters of mislabeled items, selecting them and trying to improve these regions adding or removing models initially selected by the automatic search procedure.

We started the interaction choosing the *Age* attribute in the selector of the data space to see it on the vertical axis of the plots, aiming to find clusters of mislabeled items with the data organized by that attribute. In the visualization of the data (see Figure 6, part 1), we have two classes (individuals that earn more than 50k per year; less than 50k per year) positioned horizontally in non-coincident areas of the scatter plot. Also, for each class we can analyze how the errors are distributed accordingly to the classification probability that the model gave to each data point because this information is mapped to the horizontal axis in both cases. For each class and its corresponding bin in the data space visualization, the errors that appear more to the left on the scatter plot are the ones with low classification probability to the predicted class. Conversely, to the right are the data points in which the model was wrong but gave high classification probability. This representation gave valuable input because wrongly classified points (dots with white stroke) with low probability are often easier to fix. They are nearer to the classification boundary, and we can interpret them as the cases in which the model was not so sure about the classification.

We identified regions with more errors in the classified data, but the overplotting did not help us to infer the density of points in these regions. To overcome this limitation, we enabled the heat map visualization on top of the scatter plots. The variation in color accordingly to the density of data points facilitated the identification of regions with clusters of errors in both classes. To decide which class to look for improvements, we went to the model space to discover if we had more good performing models for one of the classes (see Figure 7). We adjusted the axes to represent in one scatter plot the models accordingly to both overall performance and the performance for class 0 (less than 50k per year). We have done the same in the other model space scatter plot, but this time selecting both the performance for class 1 and overall performance. We discovered that class 1 had more available models to improve this class without hurting the overall performance (Figure 7, the cluster of models circled in red on the right image). So, we decided to try improving class 1.

Then, we selected the most dense regions in the heat map to look for greater impact w.r.t. the number of mislabeled data points that we were aiming to fix. We adjusted one of the model space scatter plots to show both the performance of the ensemble to the current data selection and the overall performance. This operation was performed by attributing to the horizontal axis the metric of performance for the current data selection, and to the vertical axis the overall performance (see Figure 6, part 2). When we set the corresponding metrics, the model space was updated to show the performance of each model with respect to the described axis selection. In the top-right area of the models scatter plot – the region that fits both conditions of local and overall good performance – we found models that were not included yet by the automatic selection procedure (see also Figure 6, part 2, in which we show where we found the models in the visualization). The automatic ensemble selection procedure is not designed to experiment all possible combinations because this is a huge combinatorial problem. So, it is always possible to find alternative combinations that were not evaluated by the automatic search.

We added the potentially good performing models for the current data selection, and we saw on the data space panel the mislabeled data points moving to the correct class (Figure 6, part 3). Our data space visualizations were designed to update themselves accordingly to any change in the model space (when the user adds or remove models). The update worked perfectly, and the selected mislabeled data items with low probability to the initially predicted class (class 2 represented in orange, Figure 6 part 1) moved to class 1 with also low classification probability, but correctly classified. With this procedure, in the provided example all the selected data changed from wrongly classified to correct classification. We confirmed in the performance text panels of our tool that the new ensemble selection did not hurt the overall performance of the initial automatic ensemble configuration. We have done that by comparing the performance of the automatically selected ensemble that always appears in our tool with the updated performance measures for the new ensemble selection.

The selectable heat map visualization we have on top of the data scatter plot helped to identify the more dense regions with classification errors in the data space. With only the scatter plots, this would not have been possible. The overplotting results in a visually similar representation of regions with a cluster of errors but does not distinguish important variations in the quantity of these errors. We performed the selection of a potentially interesting region to fix classification problems, also backed up by the linked text table with the raw data that gives the information about which range of ages, in this particular example, is part of the selection.

In the presented example, we had models available in the library to improve local regions of the data space, and by including them we achieved our goal of successfully exploring alternative ensemble configurations in the available collection of models. The data and ensemble updates (Figure 2), a core component of our approach, were fundamental to enable the sequence of described interactions. The selection of a region of interest in the data space triggered a data selection update, which allowed the representation in the model space of the classifiers' individual performance accordingly this local selection. Then, the inclusion of potentially good models to fix local classification errors triggered again an ensemble update, which updated the data panel and showed the new classification outputs for the current ensemble selection. Interacting with our the tool, the user can see in the data space the points moving horizontally because the classification probabilities change after each ensemble update. The more noticeable this movement, the more evident is to the user the impact of introducing changes in the model space.

## 5 DISCUSSION AND FUTURE WORK

In this section, we emphasize the main strengths of our work and indicate directions for future research on the integration of classification model and data spaces.

**Data vs. Feature-Space**  Many related works that use visualization for the inspection, attribute selection (also called *feature selection*), or in general the improvement of a part of the classification problem exist [6, 17, 25]. Those techniques stay at attribute level and allow the exploration of the attribute space with respect of the importance or the added value of an attribute. Typically, the goal is to adjust the attribute set, and in consequence the re-training of classification models, which is fundamentally different to the model selection task we are aiming at. We provide the complete data space in two-dimensional scatter plots, as illustrated in Figure 2, with focus on the exploration of classification errors, as these are the starting point for further adoptions of the ensemble. Compared to existing work, this effectively reduces the abstraction between the data input and the classification problem, as we omit the attribute extraction and the corresponding data transformation. Instead, we allow the user to directly work with the data that is subject to the classification, which fosters reasoning and enables findings on data record level, which is the key feature of our work.

**Interactive Exploration**  A core part of our contribution is the manual selection of regions of the data space, which is a task that can be automated, for example, by utilizing an interestingness measure. However, the search space enumeration is a very costly operation, in terms of computing time and the required computing power, as there are many different sets of data points to enumerate. To overcome this problem, we present the user scatter plot visualizations, where visual patterns created by the point positions as well as their visual mapping, guide the user to interesting areas. Additionally, the user can bring in expert or domain knowledge about the data to make informed guesses of interesting local regions as a starting point for further examination and exploration. When it comes to interactive model selection, i.e. the adaption of the classifier ensemble, we protect the performance of the ensemble by the constraint that a selection of a different model must not worsen the global performance. With this restriction, we do not require the user to understand all model differences at all, but still enable the interactive adaption of classifier ensembles. For future work, we are interested in softening this constraint, for example, by automated what-if analysis based on the current model selection. In consequence, the user can select a model that performs better locally, but also contributes to a better global performance by leaving out other models from the model ensemble, that are currently not under investigation.

**Visualization**  As described in Section 3.1, we visualize the data and model spaces using scatter plot visualizations, where each point represents a data record, or a classification model from the classifier ensemble, respectively. To get an overview of the classification outputs, the user can generate the two-dimensional data space scatter plots by applying state of the art dimensionality reduction techniques. Users can choose between different techniques to reveal linear (PCA [13], MDS [10]) or non-linear patterns (t-SNE [24]) in the data points or models. We also provide representations without dimensionality reduction as depicted, for example, in Figure 6. The classification results are binned per attribute and class and plotted horizontally accordingly to the classification probabilities of each data point, increasing the chance of finding local patterns. Scatter plots are prone for overplotting, which gives potentially wrong impressions of the data distribution, as it is nearly impossible to perceive the number of overplotted data points correctly. To cope with that issue, we integrated a heat map overlay, which displays the density of data points using a continuous interpolation, mapping point density to colors ranging from black over red and yellow to white. Alternatives to this approach, such as scatter plot matrices (SPLOMs) are available, although, they do not support the idea of an integral data space visualization. Instead, they display pairwise attribute combinations, which are subsets of the data space. Similarly, small multiples or glyph-like settings are possible, but still, it has to be decided what information is shown by the visualization, as well as how to order them meaningfully. Because the two-dimensional position of the points in the scatter plots indicates their position in the data space, we utilize in the heat map the color of the data points to indicate errors, as we are especially interested in data points that are classified wrongly.

**Generalization**  Our approach is clearly suited for wider application beyond the use that we illustrated in Section 4. Formally, our proposed workflow is comprised of the classification problem, the input data and a collection of classifiers. The classification problem can be a binary or multi-class problem, which makes the workflow applicable to all kinds of classification problems. The collection of classifiers is not restricted to Multiple Classifier Systems. Random forests, or more general, any other hybrid information system, is also suitable for our approach. The view of *Active Learning*, in which uncertainty and diversity measures are considered in prioritizing data for labeling, also seems extensible to our workflow. This view could improve the criteria for organizing the data space and potentially also improve the interactive ensemble model building process, targeting greater model performance generalization. Additionally, we also support varying parameter spaces, varying model families and arbitrary combinations of them. In consequence, our workflow is not only suitable for the selection of model families, as we demonstrated in this paper, but also for parameter space exploration. An issue of classic feature-based visualizations of classification problems is scalability. Visually, we already introduced density-based heat maps as a counter-measure to be able to scale to large data sets or model spaces. Therefore, limitations are imposed only by the available computing power, and in consequence the ability to support interactions in the model and data space. For future work, we plan to formalize our workflow and argue about its general applicability by examining related approaches, based on our formalization.

**Data overfitting**  Ensembles of classifiers can often deliver superior performance than single classifier models, but this advantage also comes with the risk of overfitting. With respect to this problem, we highlight that our approach supports both adding and removing models. It is not only an additive approach, and by removing models we provide a way to manually avoid overfitting. In any case, despite the fact that our methods can deal with data overfitting, we did not focus on model performance in this work. As future work, a more extensive performance evaluation should come together with new use cases to test the robustness under varying circumstances and validate the scope of applicability of our proposed methods.

## 6 CONCLUSION

We foster the use of visual methods for exploring model and data spaces in classification problems. Our integrative approach enables a feedback loop that keeps the user always in control of any model selection change introduced in ensembles of classifiers. We used Multiple Classifier Systems to instantiate our ideas and explore those abstract spaces. However, we can generalize and extend our workflow to any type of classifier models, combined in ensembles or not, what gives plenty of opportunities for visualization research on correlated topics.

## ACKNOWLEDGMENTS

This paper has been supported by the German Research Foundation DFG within the project *Knowledge Generation in Visual Analytics* (FP 566/17), and by the CNPq, Conselho Nacional de Desenvolvimento Cientifico e Tecnologico - Brazil.